\documentclass[aps,prb,twocolumn,showpacs,superscriptaddress]{revtex4-1}

\usepackage{natbib}
\usepackage{amssymb}
\usepackage{amsmath}
\usepackage{graphicx}
\usepackage{dcolumn}
\usepackage{bm}
\usepackage{color}

\begin{document}

\title{First principles investigation of magnetocrystalline anisotropy at the $L2_1$ Full Heusler$|$MgO interfaces and tunnel junctions}

\date{\today}

\author{Rajasekarakumar  Vadapoo}
\altaffiliation{Present address: Geophysical Laboratory, Carnegie Institution of Washington, Washington, DC 20015, USA}
\author{ Ali Hallal}
\author{Hongxin Yang}
\author{Mairbek Chshiev}
\affiliation{Univ. Grenoble Alpes, INAC-SPINTEC, F-38000 Grenoble, France}
\affiliation{CNRS, SPINTEC, F-38000 Grenoble, France} 
\affiliation{CEA, INAC-SPINTEC, F-38000 Grenoble, France}

\begin{abstract}
Magnetocrystalline anisotropy  at Heusler alloy$|$MgO interfaces have been studied using first principles calculations. 
It is found that Co terminated Co$_{2}$FeAl$|$MgO interfaces show perpendicular magnetic anisotropy up to 1.31 mJ/m$^2$, 
while those with FeAl termination exhibit in-plane magnetic anisotropy. Layer resolved analysis indicates that the origin of 
perpendicular magnetic anisotropy in Co$_{2}$FeAl$|$MgO interfaces can be attributed to the out-of-plane orbital contributions 
of interfacial Co atoms. At the same time, Co$_{2}$MnGe and Co$_{2}$MnSi interfaced with MgO tend to favor in-plane magnetic anisotropy
for all terminations. 
\end{abstract}

\pacs{75.30.Gw, 75.70.Cn, 75.70.Tj, 72.25.Mk}
\maketitle

\section*{Introduction}
Perpendicular magnetic anisotropy (PMA) in transition metal$|$insulator interfaces has been demonstrated more than a decade ago.~\cite{Monso2002,Rodmacq2003} 
These interfaces have become a viable alternative to PMA in fully metallic structures based on heavy non-magnetic elements with strong spin-orbit coupling 
(SOC)~\cite{Nakajima1998, Carcia1985, Draaisma1987, Weller1994}. Indeed, high PMA values were observed in Co(Fe)$|$MOx (M=Ta, Mg, Al, Ru, etc.) interfaces 
despite their weak SOC~\cite{Monso2002, Rodmacq2003}. These structures serve as main constituents for perpendicular magnetic tunnel junctions (p-MTJ) which 
are very promising for realizing next generation of high density non volatile memories and logic devices~\cite{Kim2008,Nistor2009,Nistor2010,Ikeda2010,Endo2010}. 
One of the most important requirements for the use of p-MTJ  in spintronic applications including high density spin transfer torque magnetic random access memory 
(STT-MRAM) is a combination of large PMA, high thermal stability and low critical current to switch magnetization of the free layer. CoFeB$|$MgO p-MTJ is one of 
the most promising candidates among state-of-the-art structures~\cite{Ikeda2010}. However, another class of ferromagnetic electrode materials with drastically 
improved characteristics for use in p-MTJ are Heusler alloys, since they possess much higher spin polarization~\cite{Graf2011} and significantly lower Gilbert 
damping~\cite{Liu2009}. 

Full Heusler alloys (X$_{2}$YZ)$|$MgO interfaces with high interfacial PMA and weak spin orbit coupling (SOC) have been gaining interest 
recently~\cite{Wu2009,Wang2010,Wen2011,Graf2011}. For instance, MgO-based MTJs with Co$_{2}$FeAl(CFA) electrodes show high PMA in most of 
the experiments. The surface anisotropy energy ($K_{s}$) is found to be around 1~mJ/m$^2$ for Pt$|$CFA$|$MgO trilayer~\cite{Li2011} and 
CFA$|$MgO~\cite{Wen2011,Cui2013} interfaces. The observed PMA values for these structures are comparable to those reported for CoFeB$|$MgO~\cite{Ikeda2010} 
and tetragonally distorted Mn$_{2.5}$Ga films grown on Cr buffered MgO~\cite{Wu2009}. However, there are reports on observation of in-plane magnetic 
anisotropy (IMA) for CFA$|$MgO interfacial structures in different cases~\cite{Belmeguenai2013,Belmeguenai2014}. Thus, these interfaces show PMA with 
values between 0.16-1.04~mJ/m$^2$~\cite{Wen2011,Okabayashi2013,Gabor2013a} as well as IMA with $K_s$=-1.8~mJ/m$^2$~\cite{Belmeguenai2013}. On the other 
hand, some theoretical studies have reported PMA values of 1.28~mJ/m$^2$ for Co terminated structures~\cite{M.TsujikawaD.Mori2013}, IMA of 
0.78~mJ/m$^2$~\cite{M.TsujikawaD.Mori2013} and PMA of 0.428~mJ/m$^2$~\cite{Bai2013a} for FeAl termination. It has been suggested that interfacial 

\begin{figure}
\centering
\includegraphics[width=1\columnwidth]{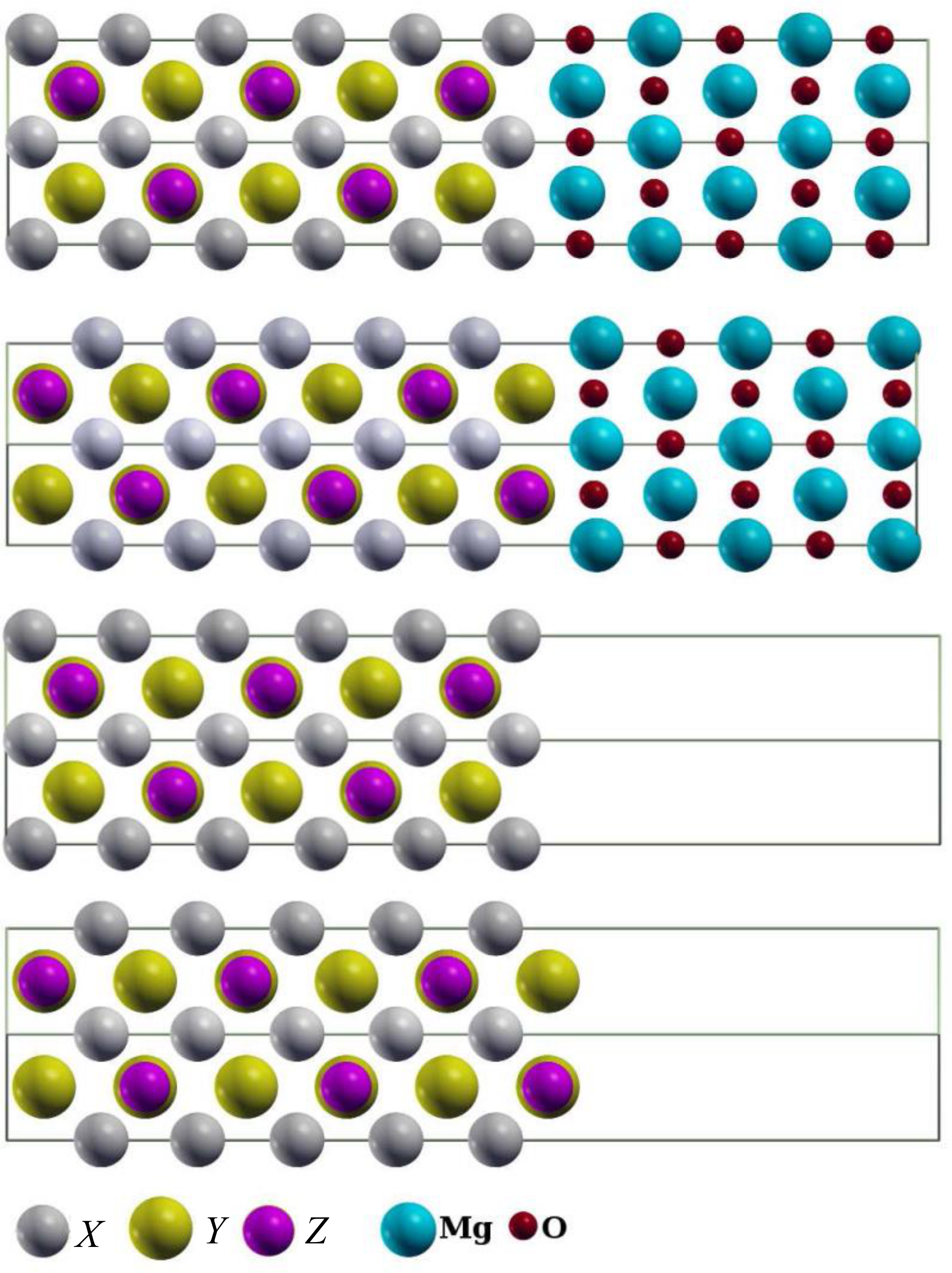} 
\caption{(Color online) Perspective view of (a) X terminated, (b) YZ terminated interface structure of Heusler (X$_{2}$YZ)$|$MgO and (c) X terminated, (d) YZ terminated
Heusler$|$Vacuum slabs with X=Co, YZ=FeAl, MnGe and MnSi. Grey, yellow, pink, blue and red spheres represent X, Y, Z, Mg and O atoms, respectively.}
\label{structure}
\end{figure}
Fe atoms are responsible for PMA in these structures~\cite{Okabayashi2013} but the microscopic origins of anisotropy remains to be clarified further.  

In order to elucidate the origin of PMA in these interfaces, we present a systematic study of magnetic anisotropy in Heusler alloy (X$_{2}$YZ)$|$MgO 
interfaces [with X=Co, YZ=FeAl, MnGe and MnSi] using first principles method. We explore the different interfacial conditions in these interfaces. In
order to understand the microscopic mechanism of PMA, we employ the onsite projected and orbital resolved analysis of magnetocrystalline anisotropy 
energy (MA) which allows identification of layer contributions along with the corresponding different orbital contributions~\cite{Yang2011,Hallal2013}. 
We found that the magnetic anisotropy is much more complex compared to that in Co(Fe)$|$MgO structures~\cite{Hallal2013} and it is strongly dependent on 
the interface termination and composition.

\section*{Methods}

Calculations are performed using Vienna ab initio siumulation package (VASP)~\cite{Kresse1993,Kresse1996} with generalized gradient approximation~\cite{Wang1991} 
and projected augmented wave potentials~\cite{Kresse1999, Blochl1994}. We used the kinetic energy cutoff of 600 eV and a Monkhorst-Pack k-point grid 
of $13 \times 13 \times 3$ where the convergence of MAE is checked with repect to the number of K-points. 
Initially the structures were relaxed in volume and shape until the force acting on each atom falls below 1~meV/\AA. The Kohn-Sham 
equations were then solved  to find the charge distribution of the ground state system without taking spin-orbit interactions (SOI) into account. Finally, the 
total energy of the system was calculated for a given orientation of magnetic moments in the presence of spin-orbit coupling using a non-self-consistent calculation. 
The surface magnetic anisotropy energy, $K_{s}$ is calculated as $(E^\parallel-E^\perp)/a^2$, where $a$ is the in-plane lattice constant and $E^\perp$($E^\parallel$) represents energy for 
out-of-plane [001](in-plane [100]) magnetization orientation with respect to the interface. The in-plane anisotropy (difference between [100] and [110]) have been checked and found to be negligible. Positive and negative values of $K_{s}$ corresponds to out-of-plane and in-plane anisotropy respectively. In addition, we define  the effective anisotropy $K_{eff} = K_{s} / t_{CFA} - E_{demag}$, where $E_{demag}$ is the demagnetization energy 
which is the sum of all the magnetostatic dipole-dipole interactions upto infinity. We adopt the dipole-dipole interaction method to calculate the $E_{demag}$ term 
instead of $2 \pi M_{s}^2$, where $M_{s}$ is the saturation magnetization; since the latter underestimates this term for thin films.~\cite{Hallal2013, Hallal2014, Schuurmans1990}
In VASP the spin-orbit term is evaluated using the second-order approximation:

\begin{equation}
H_{SOC} = \frac{1}{2(m_e C)^2} \frac{1}{r}\frac{dV}{dr} \vec{L} . \vec{s} 
\end{equation}

where $V$ denotes the spherical part of all-electron Kohn-Sham potential inside the PAW spheres, while $\vec{L}$ and $\vec{s}$ represent the angular momentum operator and the Pauli spin matrices, respectively. The spin-orbit coupling then can be calculated for each orbital angular momentum, and from which one can extract layer- and orbital-resolved MAE~\cite{Hallal2013, Hallal2014, Peng2015, Yang2015}.

\begin{figure}
\centering
\includegraphics[width=0.4\textwidth ]{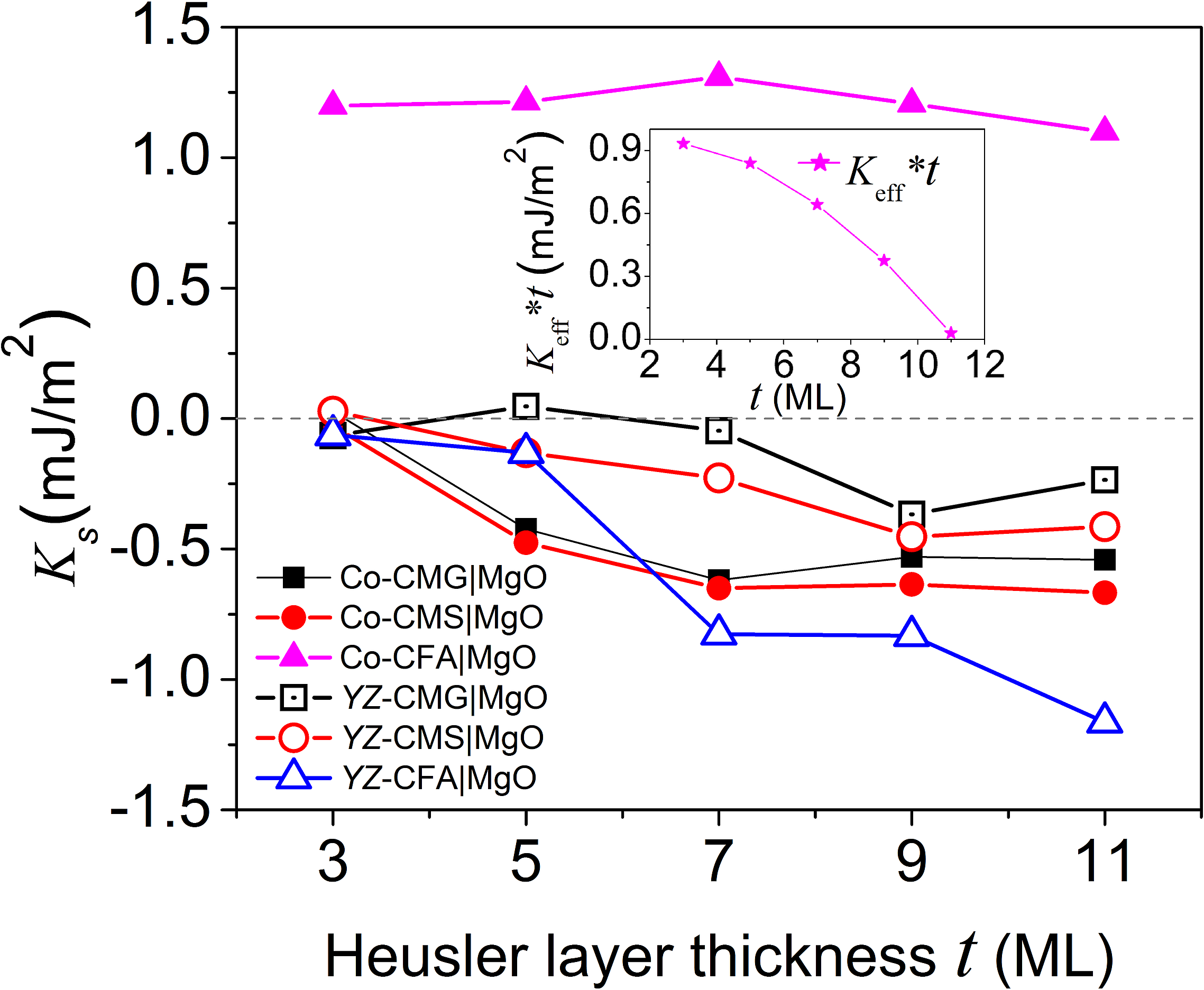} 
\caption{(Color online)  Surface magnetic anisotropy energy ($K_{s}$) as a function of number of heusler atomic-layers (ML) in Co and YZ terminated heusler 
(X$_{2}$YZ)$|$MgO structures. Filled data points represent Co terminated and open data points represent YZ terminated interfaces.  Blue triangle represent 
Co$_{2}$FeAl (CFA), black square  for Co$_{2}$MnGe (CMG) and red circle for Co$_{2}$MnSi (CMS) interfaces. Inset shows the effective anisotropy ($K_{eff}*t$) 
as a function of thickness of CFA in Co terminated CFA$|$MgO interface.}
\label{PMA}
\end{figure}

\section*{Results}
Full-heusler $(X_{2}YZ)$ alloys are  intermetallic compounds with cubic $L2_{1}$ structure and belongs to the space group $Fm\overline{3}m$~\cite{Graf2011,Culbert2008}. 
The magnetocrystalline anisotropy of bulk heusler is found to be  negligible. The Heusler$|$MgO interfaces have been setup with the  crystallographic orientation of 
Heusler(001)[100]$\parallel$MgO (001)[110]~\cite{Yamamoto2006,Gabor2011,Grunebohm2009,Bai2013a}. This results in a relatively low lattice mismatch between Heusler(001) 
and MgO(001) with a 45 degrees in-plane rotation. The energetically stable X and YZ terminations at the interface were studied and will be denoted as X-Heusler$|$MgO 
and YZ-Heusler$|$MgO as shown respectively in Fig.~\ref{structure}(a) and (b). The results of these interfaces will be compared to those of X-Heusler$|$Vacuum and 
YZ-Heusler$|$Vacuum slabs shown in Fig.~\ref{structure}(c) and (d), respectively. 

\begin{figure}
\centering
\includegraphics[width=0.4\textwidth ]{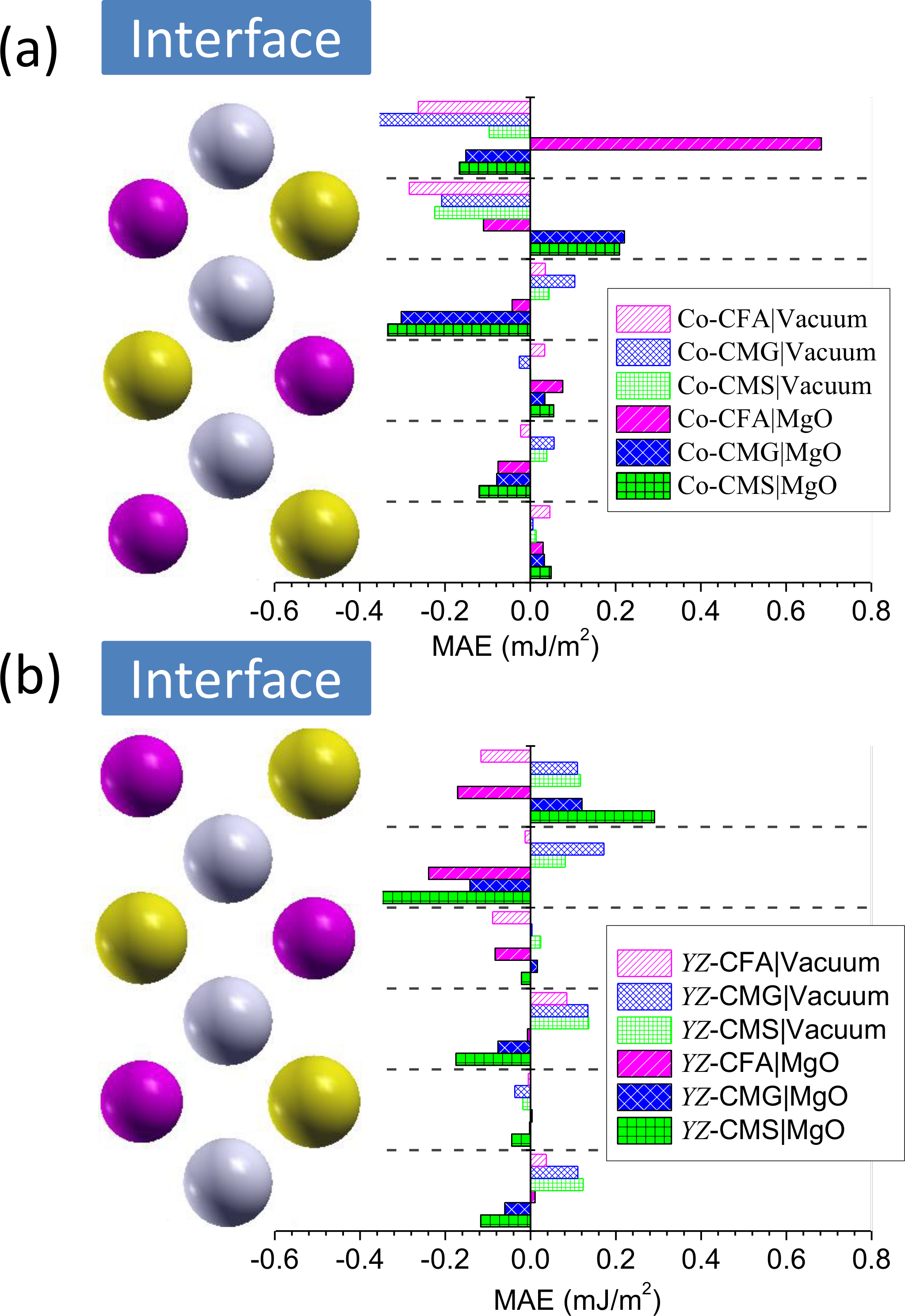} 
\caption{ (Color online) Atomic layer resolved contributions to the anisotropy for (a) X - terminated and (b) YZ - terminated Heusler$\vert $MgO  (solid filled bars) and Heusler$ \vert $vacuum (light filled bars) structures shown in Fig.~\ref{structure}(a,c) and (b,d), respectively.  The Co$_{2}$FeAl (CFA), Co$_{2}$MnGe (CMG) and Co$_{2}$MnSi (CMS) cases are represented by pink, blue and green bars, respectively. The side view of the corresponding Heusler layer are shown on the left for convenience.}
\label{layer}
\end{figure}

Increasing the MgO thickness beyond 5 atomic-layers (ML) is found to have no effect on magnetic anisotropy. The variation of suface magnetic anisotropic energy 
($K_{s}$) with the thickness of Heusler layers varying from 3 to 11 ML for the Heusler$|$MgO interfaces is shown in Fig.~\ref{PMA}. One can see that only 
Co-CFA$|$MgO structure gives rise to very high PMA which weakly depend on CFA thickness, while the FeAl-CFA$|$MgO and all CMG$|$MgO as well as CMS$|$MgO 
show IMA. It is interesting to note that the magnetic anisotropy energy for the CMG$|$MgO and CMS$|$MgO as a function of thickness follow similar trend 
which might be due to the inert nature of Z-element (Ge, Si). The in-plane anisotropy contribution in these structures increases as a function of thickness 
and stabilizes after 9 ML.  It can be seen that $K_{s}$ for Co-CFA$|$MgO increases from $1.20$~mJ/m$^2$ to a maximum of $1.31$~mJ/m$^2$ at 7 ML thickness 
($\sim 0.8$~nm), which is in agreement with experimental findings of M.~S.~Gabor {\it et al}.~\cite{Gabor2013a} and Z.~Wen {\it et al}.~\cite{Wen2011}. Inset in 
Fig.~\ref{PMA} shows the corresponding effective anisotropy ($K_{eff}$ $\ast t$) as a function of CFA thickness ($t_{CFA}$). It  shows a decaying 
behavior and vanishes around 11 ML  becoming  IMA beyond this thickness, in  reasonable agreement with recent experiments~\cite{Wen2011,Gabor2013a}.   

\begin{figure}
\centering
\includegraphics [ width=0.4\textwidth ]{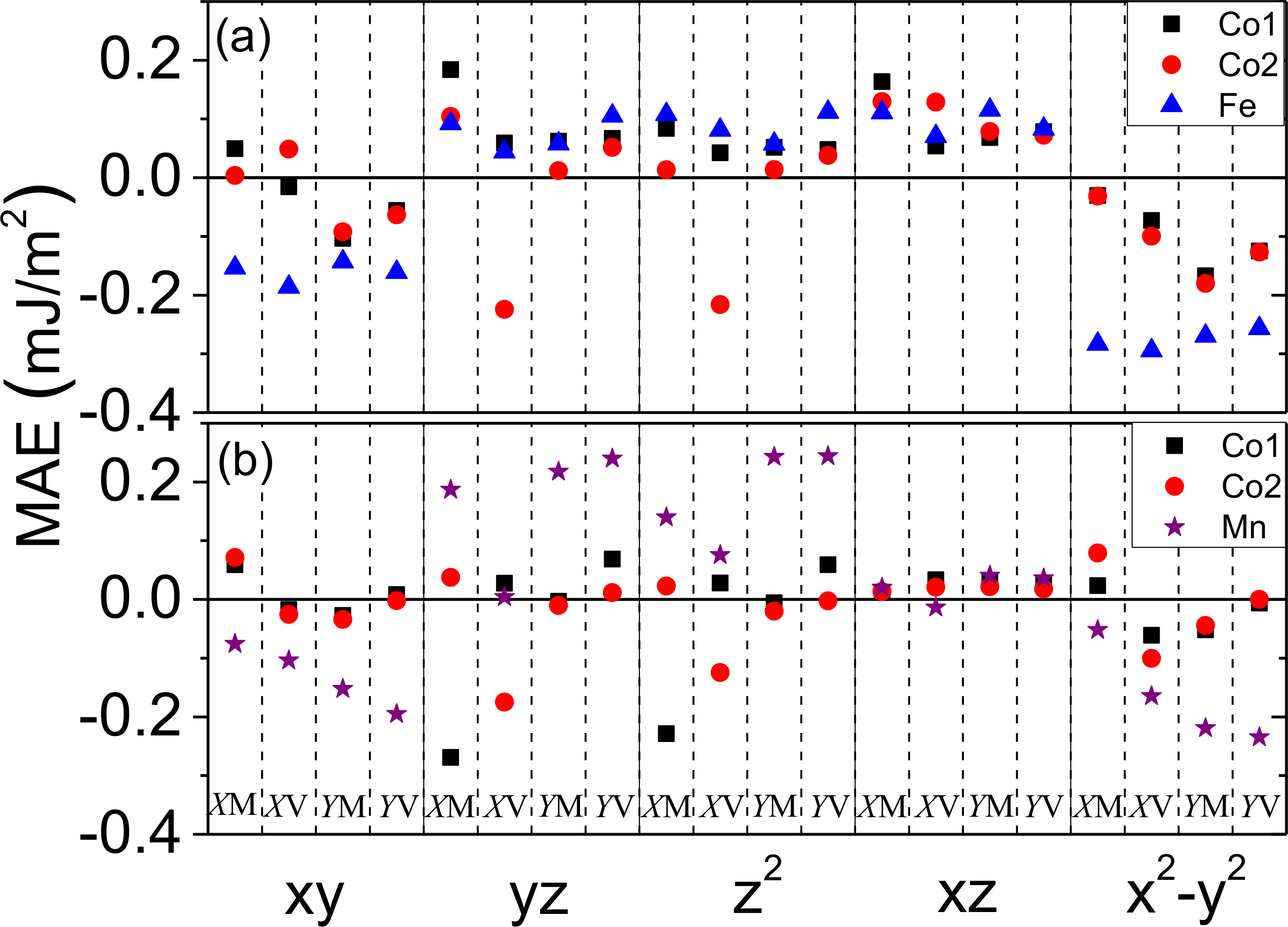}
 \caption{ (Color online) $d$-orbital resolved contributions to magnetic anisotropy for interfacial atoms in X- and YZ-terminated (a) Co$_{2}$FeAl$|$MgO  and (b) Co$_{2}$MnGe$|$MgO structures along with their free surface counterparts. Black square, red circle, blue triangle and purple star represent contributions from orbitals with $d$-character of two Co (Co1 and Co2 within the same atomic-layer), Fe and Mn atoms, respectively. XV, XM, YV and YM denote X-Heusler$|$Vacuum, X-Heusler$|$MgO, YZ-Heusler$|$Vacuum and YZ-Heusler$|$MgO interfaces respectively.}
 \label{orbital}
\end{figure}

\begin{figure*}
\centering
\includegraphics [ width=0.9\textwidth ]{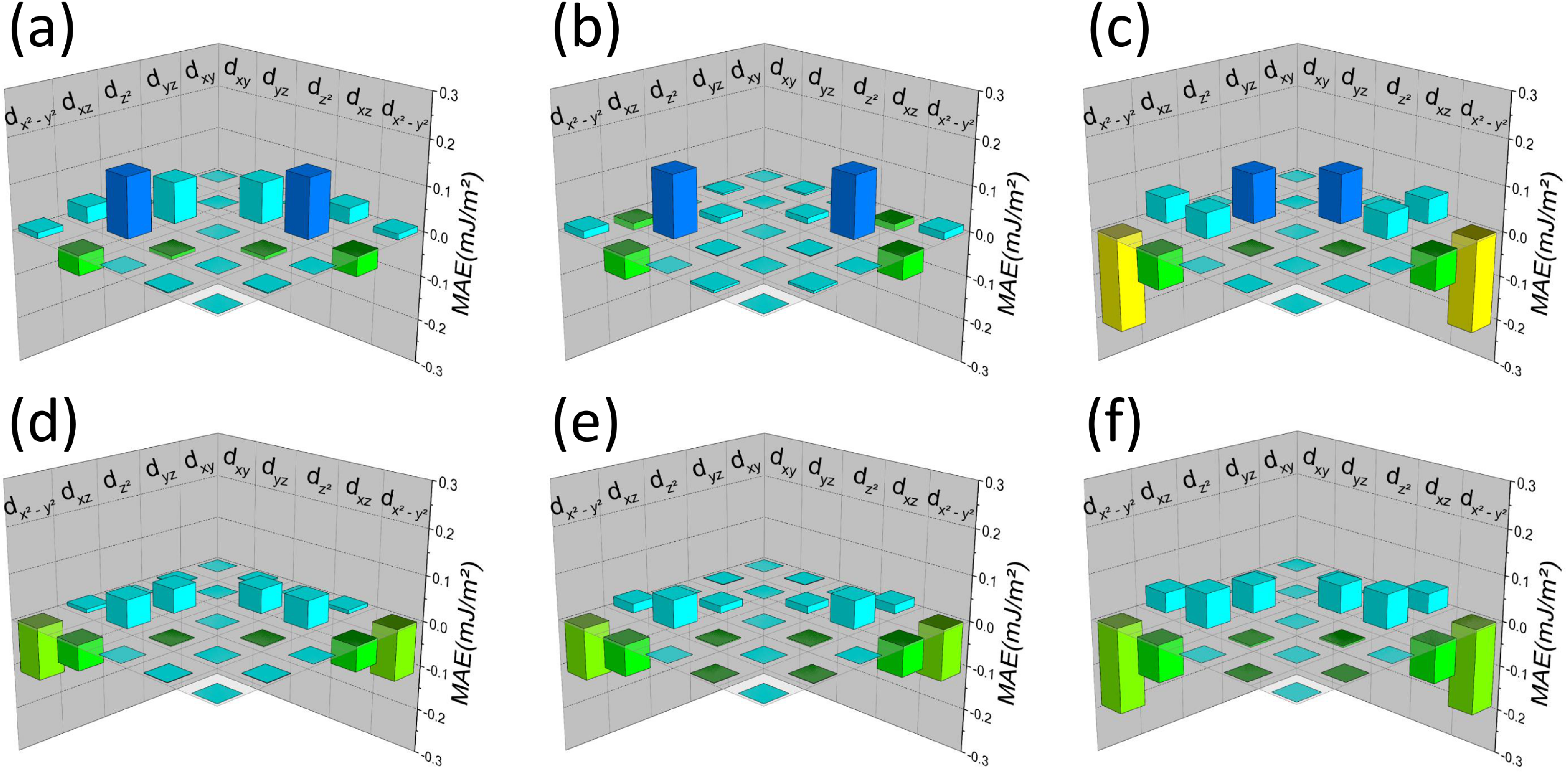}
 \caption{ (Color online) Magnetic anisotropy contribution from different  d-orbital hybridizations at the interfacial atoms of (a)[(d)] Co1  (b)[(e)] Co2 and (c)[(f)] Fe   for Co-terminated Co$_{2}$FeAl$|$MgO interface [ FeAl-terminated interface].}
 \label{orbitalmatr}
\end{figure*}
In order to understand the origin of PMA and effect of MgO, we examined the on-site projected magnetic anisotropy for the 11 ML of Heusler$|$MgO and their 
free surface counterparts as shown in Fig.~\ref{layer}. As one can see, the major PMA contribution of 0.69 $mJ/m^2$ in Co-CFA$|$MgO structure comes from the 
interfacial Co atoms while the inner layers show fair amount of in-plane or out-of-plane contributions represented by solid pink bars in~Fig.~\ref{layer}(a). 
By comparing with CFA$|$Vacuum shown by unfilled pink bars in the same figure, we can clearly identify that the presence of MgO on top of Co layer plays a 
decisive role in establishing the PMA in Co terminated CFA$|$MgO structure. More complicated behavior is observed for Co-CMG and Co-CMS structures where the 
role of MgO in anisotropy varies depending on layer. While it tends to decrease(increase) the IMA in the 1st Co layer for Co-CMG(Co-CMS), it simultaneously 
flips the IMA into PMA (PMA into IMA) for 2nd YZ (3rd Co) layer.  

Similar nontrivial picture is observed for YZ terminated structures shown in Fig.~\ref{layer}(b). By employing the same analysis in 
order to clarify the role of MgO vs vacuum next to YZ-terminated Heusler alloy, one can see that the MgO has a tendency to improve the IMA for 
the case of YZ-CFA for all layers. Furthermore, it enhances the PMA(IMA) for the 1st(all Co) layers of YZ-CMS and YZ-CMG structures. 

Overall it can be concluded that the presence of MgO tends to favor IMA from all Co layers except the interfacial ones in Co-CFA and Co-CMG structures. 
At the same time, the inner YZ layers in presence of MgO have a tendency for the PMA for Co-terminated structures (Fig.~\ref{layer}(a)), 
while YZ interfacial layer favor the IMA(PMA) in YZ-CFA(YZ-CMS and YZ-CMG)~(cf.~Fig.~\ref{layer}(b)).

To further elucidate the microscopic origin of PMA, we carried out the $d$-orbital resolved magnetic anisotropy contributions for 
interfacial atoms as shown in Fig.~\ref{orbital}. One can see that the switch from IMA to PMA when MgO is placed on top of Co 
terminated CFA mainly arises from the out-of-plane orbitals ($d_{xz,yz}$ and $d_{z^2}$) as shown by comparison of XV and XM 
columns in Fig.~\ref{orbital}(a). Furthermore, this switch is assisted by all $d$ orbitals within the 2nd (YZ) layer. At the 
same time, the MgO-induced enhancement of the IMA in the first two layers from interface (FeAl and Co) in case of YZ terminated 
CFA (see Fig.~\ref{layer}(b)) is due to increase(decrease) of IMA(PMA) contribution from $d_{x^2-y^2}$($d_{yz}$ and $d_{z^2}$) 
orbitals, as seen from comparison of columns YV and YM in Fig.~\ref{orbital}(a). 

In the case of Co terminated CMG, the effect of MgO results in overall tendency to decrease the IMA with participation of in-plane $d$ 
orbitals ($d_{xy}$ and $d_{x^{2}-y^{2}}$) in the first Co layer with a quite interesting opposing contributions from out-of-plane $d_{yz}$ 
and $d_{z^2}$ orbitals (see XV and XM columns in Fig.~\ref{orbital}(b)). At the same time, for the second layer (MnGe) contribution, the 
presence of MgO has a clear tendency to switch from IMA into PMA assisted by all $d$-orbitals. As for Mn terminated CMG, the presence of 
MgO has almost no effect on 1st MnGe layer anisotropy contributions, while it induces the flip from PMA to IMA from almost all $d$ orbitals
within the second Co layer~(Fig.~\ref{orbital}(b)). The orbital contributions for CMS are found to be very similar to CMG orbital contributions.  

To further elucidate the PMA origin in Co$_{2}$FeAl$|$MgO, Fig.~\ref{orbitalmatr} shows magnetic anisotropy contribution originated from the spin orbit coupling induced hybridizations between different orbital channels for interfacial atoms at the Co-terminated and FeAl-terminated interface.  In all cases out-of-plane orbitals (d$_{xz(yz)}$, d$_{z^2}$) mututal hybridizations strongly favor PMA contribution. At the same time, hybridization among in-plane orbitals (d$_{xy}$, d$_{x^2-y^2}$) gives rise to IMA except for the case of Co1 and Co2 atoms in the Co-terminated interface where they have a slight PMA contribution. In all cases d$_{x^2-y^2}$ hybridization with out-of-plane, mainly d$_{yz}$, orbitals contribute to IMA. On the other hand, d$_{xy}$ hybridization with out-of-plane, mainly d$_{xz}$, orbitals contribute to PMA except for the case of Co2 atom in Co-terminated structure. However, the sum of the contribution coming from Co1 and Co2  atoms  is favoring PMA.

\section*{Discussion}

Aforementioned analysis shows that Co-CFA$|$MgO structure favors the high PMA while YZ termination in CFA$|$MgO structure 
give rise to IMA. This allows us to conclude that interfacial Co atoms are responsible for the PMA. However, it was claimed
recently that the origin of PMA could be attributed to Fe atoms at the interface in CFA$|$MgO~\cite{Okabayashi2013} using 
XMCD measurements in combination with Bruno's model analysis~\cite{Bruno1989}. In order to resolve this disagreement, we 
carried out the orbital momentum calculations for 7~ML structure corresponding to that reported in experiments~\cite{Okabayashi2013}
with both terminations. We systematically found that per layer resolved orbital moment anisotropy (OMA) is inconsistent with layer 
resolved MA contributions for Co layers while it remains in qualitative agreement for layers containing Fe. We can therefore conclude 
that Bruno's model should be used with caution and its validity may depend on particular system.

In summary, using first principles calculation we investigated the magnetic anisotropy of Full Heusler$|$MgO interfaces and MTJs for 
all terminations. It is found that Co terminated CFA$|$MgO shows the PMA of $1.31$mJ/m$^2$ induced by the  presence of MgO in agreement
with recent experiments while FeAl terminated CFA and other structures possess the IMA. We also unveiled the microscopic mechanisms of 
PMA in Heusler$|$MgO structures by evaluating the onsite projected and orbital resolved contributions to magnetic anisotropy and found 
that interfacial Co atoms are responsible for high PMA (IMA) in CFA (CMG,CMS). Finally, out-of-plane (in-plane) orbitals tend to favor mainly PMA (IMA).

\section*{Acknowledgements}

The authors thanks C. Tiusan for fruitful discussions. This work was partly supported by Eu M-era.net HeuMem, French ANR SPINHALL and SOspin projects.



\end{document}